\documentclass[a4paper,12pt, epsfig]{article}
\def\letter{0}\def\pr{0}
\pdfoutput=1 
\usepackage{epsfig}
\usepackage{epstopdf}
\usepackage{graphicx}
\usepackage{ifthen}
\usepackage[numbers,sort&compress]{natbib}
\usepackage{ulem}

\usepackage{amsmath}
\usepackage[psamsfonts]{amssymb}
\usepackage{euscript}

\usepackage{latexsym}
\usepackage[arrow,matrix,curve]{xy}

\usepackage[hypertexnames=false]{hyperref}
\hypersetup{
    colorlinks=true,
    linkcolor=blue,
    filecolor=magenta,   
    citecolor=black,   
    urlcolor=cyan,
    }

\newskip\humongous \humongous=0pt plus 1000pt minus 1000pt

\newif\ifdtup

\def\,{\hspace{-.1cm}}
\def\hsp{,\hspace{.7cm}}

\def\fc#1#2 {\frac{n}{q}#1\frac{n}{q}#2}

\newcommand{\vac}{\ensuremath{|0\rangle}}

\renewcommand{\tanh}{\textrm{tanh}}
\newcommand{\sech}{\textrm{sech}}
\newcommand{\csch}{\textrm{csch}}

\def\exp#1{\hbox{\rm exp}\left[#1\right]}

\renewcommand{\theequation}{\arabic{section}.\arabic{equation}}
\renewcommand{\(}{\begin{equation}}
\renewcommand{\)}{end{equation} \vspace{-.05in}\linebreak}

\newcounter{saveeqn}
\newcounter{savealpheqn}

\newcommand{\alpheqn}{\setcounter{saveeqn}{\value{equation}}%
  \stepcounter{saveeqn}\setcounter{equation}{0}%
  \renewcommand{\theequation}{\mbox{\arabic{section}.\arabic{saveeqn}
\alph{equation}}}
  \renewcommand{\)}{\end{equation}}}
\def\part#1{\frac{\partial}{\partial{#1}}}%
\def\group#1{\refstepcounter{equation}\setcounter{saveeqn}
 {\value{equation}}%
  \label{#1}\setcounter{equation}{0}%
\renewcommand{\theequation}{\mbox{\arabic{section}.\arabic{saveeqn}
\alph{equation}}}
  \renewcommand{\)}{\end{equation}}}
\newcommand{\reseteqn}{\setcounter{equation}{\value{saveeqn}}%
  \renewcommand{\theequation}{\arabic{section}.\arabic{equation}}%
  \renewcommand{\)}{\end{equation}}}

\newcommand{\aalpheqn}{\setcounter{saveeqn}{\value{equation}}%
  \stepcounter{saveeqn}\setcounter{equation}{0}%
  \renewcommand{\theequation}{\mbox{
        \Alph{subsection}.\arabic{saveeqn}\alph{equation}}}
   \renewcommand{\)}{\end{equation}}}
\newcommand{\areseteqn}{\setcounter{equation}{\value{saveeqn}}%
  \renewcommand{\theequation}{\Alph{subsection}.\arabic{equation}}%
  \renewcommand{\)}{\end{equation}}}

\renewcommand{\thefootnote}{\alph{footnote}}
\renewcommand{\(}{\begin{equation}}
\renewcommand{\)}{\end{equation}}
\newcommand{\ba}{\begin{eqnarray}}
\newcommand{\ea}{\end{eqnarray}}
\newcommand{\cbp}{\mathop{\vtop{\ialign{##\crcr
   $\hfil\displaystyle{}\hfil$\crcr\noalign{\kern-13pt\nointerlineskip}
   \BIG{)}\hskip0pt\crcr\noalign{\kern3pt}}}}}
\newcommand{\pa}{\mathop{\vtop{\ialign{##\crcr

$\hfil\displaystyle{\oplus}\hfil$\crcr\noalign{\kern+1pt\nointerlineskip
}
   \hspace{.08in}$^{\alpha=0}$\hskip6pt\crcr\noalign{\kern3pt}}}}}
\renewcommand{\hsp}{,\hspace{.3in}}
\newcommand{\p}{^\prime}

\catcode`\@=11
\def\vereq#1#2{\lower3pt\vbox{\baselineskip1.5pt \lineskip1.5pt
\ialign{$\m@th#1\hfill##\hfil$\crcr#2\crcr\sim\crcr}}}
\catcode`\@=12

\renewcommand{\(}{\begin{equation}}
\renewcommand{\)}{\end{equation}}

\def\pin#1{\int \frac{d#1}{2\pi}}

\def\df{\mathcal{D}_{f}}

\def\I{\mathcal{I}}

\newcommand{\beas}{\begin{eqnarray*}}
\newcommand{\eeas}{\end{eqnarray*}}

\newcommand{\bquo}{\begin{quote}}
\newcommand{\enqu}{\end{quote}}

\def\lim#1{\stackrel{\rm{lim}}{{}_{#1}}}

\newcommand{\g}{\mathfrak g}

\def\ok#1{\omega_{k_{#1}}}
\def\okp#1{\omega_{k\p_{#1}}}

\newcommand{\beq}{\begin{equation}}
\newcommand{\eeq}{\end{equation}}
\newcommand{\bea}{\begin{eqnarray}}
\newcommand{\eea}{\end{eqnarray}}
\newcommand{\bal}{\begin{align}}
\newcommand{\eal}{\end{align}}

\newskip\humongous \humongous=0pt plus 1000pt minus 1000pt

\newif\ifdtup

\catcode`\@=11

\ifthenelse{\equal{\letter}{0}}{ 

\@addtoreset{equation}{section}
\def\theequation{\arabic{section}.\arabic{equation}}

\def\@normalsize{\@setsize\normalsize{15pt}\xiipt\@xiipt
\abovedisplayskip 14pt plus3pt minus3pt%
\belowdisplayskip \abovedisplayskip
\abovedisplayshortskip \z@ plus3pt%
\belowdisplayshortskip 7pt plus3.5pt minus0pt}

\def\small{\@setsize\small{13.6pt}\xipt\@xipt
\abovedisplayskip 13pt plus3pt minus3pt%
\belowdisplayskip \abovedisplayskip
\abovedisplayshortskip \z@ plus3pt%
\belowdisplayshortskip 7pt plus3.5pt minus0pt
\def\@listi{\parsep 4.5pt plus 2pt minus 1pt
      \itemsep \parsep
      \topsep 9pt plus 3pt minus 3pt}}

\relax

\def\section{\@startsection{section}{1}{\z@}{3.5ex plus 1ex minus  .2ex}{2.3ex plus .2ex}{\large\bf}}

\def\thesection{\arabic{section}}
\def\thesubsection{\arabic{section}.\arabic{subsection}}

\def\appendix{\setcounter{section}{0}
 \def\thesection{Appendix \Alph{section}}
 \def\thesubsection{\Alph{section}.\arabic{subsection}}
 \def\theequation{\Alph{section}.\arabic{equation}}}
\renewcommand{\theequation}{\arabic{section}.\arabic{equation}}

}{
\renewcommand{\theequation}{\arabic{equation}}

} 

\begin{document}
\def\thefootnote{\fnsymbol{footnote}}

\def\thetitle{Elastic Kink-Meson Scattering in the $\Phi^4$ Double-Well Model}
\def\autone{Jarah Evslin}
\def\auttwo{Kehinde Ogundipe}
\def\autthree{Bilguun Bayarsaikhan}

\def\affa{Alikhanian National Laboratory (Yerevan Physics Institute), Alikhanian Brothers Street 2, Yerevan 0036, Armenia}
\def\affb{MOE Key Laboratory for Nonequilibrium Synthesis and Modulation of Condensed Matter, \\
School of Physics, Xi’an Jiaotong University, Xi’an 710049, China}
\def\affc{Institute of Modern Physics, NanChangLu 509, Lanzhou 730000, China}
\def\affd{University of the Chinese Academy of Sciences, YuQuanLu 19A, Beijing 100049, China}

\ifthenelse{\equal{\pr}{1}}{
\title{\thetitle}
\author{\autone}
\author{\auttwo}
\author{\autthree}
\affiliation {\affa}
\affiliation {\affb}
\affiliation {\affc}
\affiliation {\affd}

}{

\begin{center}
{\large {\bf \thetitle}}

\bigskip

\bigskip


{\large \noindent  
\auttwo{${}^{1, 2, 3, 4}$}
\footnote{kehindeoogundipe@gmail.com}
\ and \autthree{${}^{3,4}$}
\footnote{ph.bilguun@gmail.com}
}

\vskip.7cm

1) \affa\\
2) \affb\\
3) \affc\\
4) \affd\\

\end{center}

}

\begin{abstract}
\noindent
We calculate the leading order amplitude and probability for the elastic scattering of an elementary meson and a kink in the $\phi^4$ double-well model.  Classically, the kink is reflectionless, and so the leading contribution arises at one loop.  At this order, the scattering amplitude exhibits a pole when the incoming meson energy is twice the shape mode energy, corresponding to the excitation of an unstable resonance with the twice excited shape mode.  We expect that higher order corrections will give this resonance a width equal to the inverse of the known lifetime of this unstable excitation.

\end{abstract}

%
\setcounter{footnote}{0}
\renewcommand{\thefootnote}{\arabic{footnote}}

\ifthenelse{\equal{\pr}{1}}
{
\maketitle
}{}

\section{Introduction}

The $\phi^4$ model, renowned for its wide applicability across diverse fields, stands as one of the most extensively studied models in theoretical physics\cite{2011tm, 2023zmc}. In the (1+1)-dimensional real scalar $\phi^4$ model, a notable feature is the presence of a topological soliton known as the kink. This object embodies a stable, particle-like field excitation and finds relevance in disciplines ranging from cosmology and condensed matter to particle physics, biology, and quantum optics\cite{1962vs, 1982ga, 2000jqa, 2002qs, 2004tk, 2006zz,  2008qp, 2010bd, 2011tm, 2011wg, 2013ju, 2013gpp, 2014uha, 2016txo, 2015bra, 2016cnk, 2017jmv, 2023zmc}. Despite its apparent simplicity, the $\phi^4$ kink captures many essential aspects of soliton dynamics, and methodologies developed in this lower-dimensional context often extend to higher-dimensional gauge theories such as Quantum Chromodynamics (QCD), motivating a thorough understanding of its scattering properties.

The first connection between solitons and particle physics was pioneered in the 1960s by Skyrme, who constructed baryon states as topological solitons (skyrmions) in Refs.~\cite{1961vq, 1962vh}. Later, Ref.~\cite{1991jf} initiated systematic studies of meson–kink scattering in scalar field theories, demonstrating how the spectral properties of the kink’s stability operator govern soliton–meson interactions. Despite this early progress, meson–kink scattering remained underexplored until more recent works highlighted its importance in nonlinear dynamics, including the role of radiative modes and shape excitations in kink interactions Refs.~\cite{2011yzp, 2018yzp, 2023ypw, 2023egm}.

A crucial benchmark for such studies is the Sine-Gordon model, whose integrability ensures that kink–meson elastic scattering amplitudes vanish at the quantum level due to delicate cancellations~\cite{2023egm}. In contrast, the $\phi^4$ model is non-integrable, and therefore elastic scattering does not vanish. Instead, its kinks display a wealth of nonlinear behaviors, such as kink bounce, resonance windows, bion formation \cite{1983xu}. One can also study kink scattering off localized impurities introduced by hand. However, such impurities are not part of the pure $\phi^4$ model, and similar impurity scattering can equally well be engineered in integrable models such as Sine-Gordon by coupling them to an external defects \cite{1992dk, 1992fpl}. 
Setting impurities aside, this sharp contrast makes the $\phi^4$ model a natural testing ground for exploring how integrability breaking manifests in soliton–meson dynamics.
The theoretical analysis of soliton–meson scattering and quantum kink dynamics is most naturally treated using semiclassical quantization methods. Within these approaches, a particularly important method constructs a quantum kink Hamiltonian using displacement operators built from the classical kink profile, implementing a unitary transformation of the regularized vacuum Hamiltonian. This formalism provides a systematic way to incorporate loop corrections and treat kink zero modes without invoking collective coordinates (that is, without introducing a dynamical kink-position variable or imposing the associated orthogonality constraints required in the traditional moduli-space approach), and has been developed and refined in Refs.~\cite{2019xte, 2021gxs, 2021gxs2}. A different approach in Ref.~\cite{2001dy} computes one-loop renormalized interface energies using scattering data and finite-energy sum rules. This approach does not employ unitary transformations and is restricted to one loop, but offers an elegant alternative based solely on spectral information. For comparison, the traditional collective coordinate method ~\cite{1974dc, 1975wt} approximates kink dynamics by promoting a small number of classical moduli to quantum degrees of freedom. Although powerful, it becomes cumbersome at higher loop order, motivating the more recent displacement operator approach.


Recent works, particularly by Evslin and collaborators, have applied this framework to both sine-Gordon and $\phi^4$ models, computing loop corrections to kink states as well as meson–kink scattering amplitudes \cite{2019xte, 2021nsi, 2022opz, 2022hfk, 2023egm}. These analyses revealed that sine-Gordon amplitudes vanish due to integrability\cite{1984dx}, whereas in the $\phi^4$ case the elastic scattering amplitude is finite and momentum-dependent, providing direct evidence of its non-integrability. This non-vanishing amplitude reflects the persistence of loop-level quantum contributions, which fail to cancel in the non-integrable theory.

In this work, our goal is to build upon this foundation by computing the elastic kink–meson scattering amplitude in the (1+1)-dimensional $\phi^4$ model. Our approach relies on the quantum displacement operator framework developed by Evslin and collaborators, supplemented by analytic perturbation decompositions and subleading state corrections that have recently been established for the $\phi^4$ kink\cite{2021nsi, 2022xdz}. Within this framework, we will demonstrate explicitly that quantum loop contributions give rise to a non-vanishing elastic amplitude in the $\phi^4$ model. This result is then contrasted with the sine-Gordon case, where the vanishing of the amplitude highlights the central role of integrability in suppressing scattering processes. In this way, we provide a quantum treatment of kink–meson scattering that not only clarifies the distinction between integrable and non-integrable theories, but also establishes a framework that can be generalized to other soliton-bearing models lacking integrability.

The structure of our paper is organised into this outline. In Section 2, we review the linearized soliton sector perturbation theory, the kink wave packet definition and the general analytical calculation of the amplitude of elastic scattering off of a reflectionless quantum kink. Section 3 will focus on the numerical calculation in the case of the phi-4 model. Finally, the conclusions are summarized in Section 4.

\section{Review}
\subsection{Linearized Kink Perturbation Theory}
We begin with a brief review of the linearized soliton perturbation theory, which is formulated in Refd.~\cite{2019xte, 2021gxs2}.
Consider a general Hamiltonian $H$ in the $1+1$ dimensional theory characterized by a scalar field $\phi(x)$ and its conjugate field $\pi(x)$ operating within the Schrodinger picture. In this framework, the Hamiltonian takes the form: 
\beq
H=\int d x: \mathcal{H}(x):_a
\eeq
with the local Hamiltonian density $\mathcal{H}$ expressed as:
\beq 
\quad \mathcal{H}(x)=\frac{\pi^2(x)}{2}+\frac{\partial_x \phi(x)\partial_x \phi(x)}{2}+\frac{1}{\lambda} V(\sqrt{\lambda} \phi(x)).
\eeq

Here, the degenerate potential $V(\sqrt{\lambda} \phi(x))$ with 2 minima 
with respect to $\phi(x)$, and an expansion parameter $\sqrt{\lambda}$ 
in the representation that refers to the perturbative expansion of the potential in powers of  $\sqrt{\lambda}$, are involved. At the classical level under nontrivial boundary conditions, the classical equation of motion admits a static kink solution $\phi(x,t) = f(x)$.

In the quantum theory, the normal ordering prescription is defined with respect to the plane wave vacuum of mass $m$, ensuring the removal of ultraviolet tadpole divergences arising from single vertex loop diagrams. The normal ordering mass is determined by the second derivative of the potential evaluated at the asymptotic kink vacuum:
\beq
m^2=V^{(2)}(\sqrt{\lambda} f(x))\Big|_{x=\pm\infty}\label{dec}.
\eeq
More generally, we define
\beq
V^{(n)}(\sqrt{\lambda} \phi(x))=\frac{\partial^n V(\sqrt{\lambda} \phi(x))}{\partial (\sqrt{\lambda} \phi(x))^n}.
\eeq
It is expected that the masses have the same value on the two sides of the kink to avoid uninteresting kink acceleration at the one-loop level\cite{wstab}.

In the defining frame, let $|K\rangle$ be the Hamiltonian eigenstate with the lowest energy in the kink sector
\beq
H|K\rangle=Q|K\rangle.
\eeq
This sector consists of states with a single kink and a finite number of mesons. The quantum excitations of $\phi(x)$ are referred to as mesons. The kink sector is created by acting with the displacement operator 
\beq 
\df=\exp{-i\int dx f(x) \pi(x)}
\eeq
on a state in the vacuum sector. The vacuum sector consists of states containing no kinks but a finite number of mesons, and $|0\rangle$ denotes its lowest energy state. Acting with $\df$ produces the lowest energy kink state,
\beq
|K\rangle = \df|0\rangle.
\eeq
The created kink sector, which appears to be non-perturbative, can be constructed 
using a passive transformation which intuitively shifts $\phi(x)$ by the classical kink profile
\beq
\df^\dag \phi(x) \df= \phi(x) + f(x).
\eeq
In particular, given a state $|\psi\rangle$ in the defining frame, we introduce the corresponding state in the kink frame by
\beq
|\psi\rangle^\prime = \df^\dag|\psi\rangle.
\eeq
If $|\psi\rangle^\prime$ is time-independent and therefore an eigenstate of $H$, then $|\psi\rangle$ is an eigenstate of the kink Hamiltonian
\beq
H^\prime = \df^\dag H \df. \label{hpdef}
\eeq

This transformation removes $\df$ from the state definitions, enabling the application of standard perturbation theory, although with the transformed Hamiltonian $H^\prime$.

The perturbative analysis in the vacuum sector studies the small fluctuations around the vacuum, classically represented by plane waves. In the kink sector, perturbation theory instead describes small fluctuations around the classical kink profile. Thus, we write 

\beq
\phi(x,t) = f(x) + e^{-i\omega t}\mathfrak{g}(x).
\eeq
The fluctuation $\mathfrak{g}(x)$ satisfies the Sturm-Liouville equation
\beq
V^{\prime\prime}\left(f(x)\right)  \mathfrak{g}(x) - \mathfrak{g}^{\prime\prime}(x) = \omega^2 \mathfrak{g}(x).
\eeq
The solutions decompose into different mode classes according to their frequencies $\omega$. The zero modes $\mathfrak{g}_{B} (x)$ have $\omega_B = 0$, while the continuum modes $\mathfrak{g}_{k}(x)$ have frequencies
\beq
\omega_k = \sqrt{m^2 + k^2}.
\eeq
Additionally, discrete real shape modes $\mathfrak{g}_{S}(x)$ exist with frequencies $0<\omega_S<m$. The completeness relations hold: 
\beq
\int dx \mathfrak{g}_{k_1}(x) \mathfrak{g}_{k_2}(x) = 2\pi \delta(k_1-k_2) \hsp
\int dx \mathfrak{g}_{S_1}(x)\mathfrak{g}_{S_2}(x) = \delta_{S_1S_2} \hsp
\int dx |\mathfrak{g}_B(x)|^2 = 1.
\label{normal}
\eeq
The sign of $\mathfrak{g}_{B}$ is conventionally fixed by
\beq
\mathfrak{g}_{B}(x) = -\frac{f^\prime (x)}{\sqrt{Q_0}},
\eeq
where $Q_0$ denotes the mass of the leading-order kink.

\subsection{Perturbative Expansion and Kink-Meson Scattering}

Following the formalism in\cite{1976im, 2019xte}, we decompose the field and its conjugate momentum in terms of normal modes. Quantization promotes the mode coefficients to annihilation and creation operators $B_S$, $B_{S}^{\dagger}$ and $B_k$, $B_{k}^{\dagger}$, obeying
\begin{equation}
    [B_{S}, B_{S}^{\dagger}] = 1, \quad [B_{k}, B_{k}^{\dagger}] = 2\pi \delta(k - k').
\label{opera}
\end{equation}
With these conventions, Eqs.~(\ref{normal}) and~(\ref{opera}), the field and conjugate momentum take the form
\begin{align}
\phi(x)
&= \phi_0 \mathfrak{g}_B(x)
+ \sum_S\frac{g_S(x)}{2\omega_S}\,(B_S^\dagger + B_S)
+ \int \frac{dk}{2\pi} \frac{g_k(x)}{2\omega_k}\,(B_k^\dagger + B_{-k}),
\tag{2.10a}
\\[6pt]
\pi(x)
&= \pi_0 \mathfrak{g}_B(x)+i\sum_S\frac{\mathfrak{g}_S(x)}{2}\,(B_S^\dagger - B_S)
 + i \int \frac{dk}{2\pi} \frac{\mathfrak{g}_k(x)}{2}\,(B_k^\dagger - B_{-k}).
\tag{2.10b}
\end{align}
The kink-sector vacuum $\vac_0$ also known as the kink ground state is defined to be annihilated by all fluctuation operators,
\begin{equation}
 B_S \vac_0 = B_k \vac_0 = 0,
\end{equation}
and an $n$-meson state is constructed as
\begin{equation}
|k_1 \dots k_n\rangle_0 = B_{k_1}^{\dagger} \dots B_{k_n}^{\dagger} \vac_0.
\end{equation}
Expanding a general kink sector state in terms of these basis elements,
\begin{equation}
|\psi\rangle = \sum_{m,n} \phi_0^m \int \frac{dk_1 \dots dk_n}{(2\pi)^n} \gamma_{\psi}^{mn}(k_1 \dots k_n) |k_1 \dots k_n\rangle_0,
\end{equation}
we extract relevant coefficients using old fashioned perturbation theory. For single-kink single-meson states, the elastic scattering amplitude is determined by the coefficient $\gamma^{mn}$. The key expression that governs this behavior is 
\begin{equation}
\gamma_{2k_1}^{01}(k_2) = -\frac{(\hat{\gamma}_{2k_1}^{21}(k_2) - \rho_{k_1}(k_2))}{\omega{k_1} - \omega_{k_2}},
\end{equation}
as derived in Ref.~\cite{2022pct} where $\hat{\gamma}$ and $\rho$ are determined by interaction potentials and mode overlaps, subject to normalization choices.

\subsection{Reflective Coefficient Calculation}
In one-dimensional nonrelativistic quantum mechanics, the reflection coefficient can be calculated by solving the time-independent Schrödinger equation with the boundary condition of no incoming particles from the right. The inner product of the Hamiltonian eigenstate is then taken with an outgoing wave packet on the left. This requires choosing an eigenstate with an energy in the continuum, which is non-normalizable, but still gives a finite and meaningful result \cite{2023egm}.
In studying a kink and considering a meson, which we might refer to as a small fluctuation, interacting with it, we need to consider the Hamiltonian eigenstate $|k_1\rangle$ whose energy eigenstates do not evolve in time except for a phase. 
To describe incoming or outgoing mesons, we need a wave packet - a superposition of these eigenstates - not a single eigenstate.

\subsubsection{Kink Wave Packets}
To construct the localized wave packet made from eigenstates $|k_1\rangle$, the idea is to superpose the eigenstates for a packet centered around the momentum $k_0$ and position $x_0$.
\beq
|t = 0\rangle = \pin{k} f(k_1) |k_1\rangle \label{pack}
\eeq
where $f(k_1)$ is the weighting function that:
\begin{itemize}
    \item centers the packet around momentum $k_0$
    \item localizes the packet around position $x_0$.
\end{itemize}
Now, we choose
\beq
f(k_1) = e^{-\sigma^2(k_1-k_0)^2} e^{-i(k_1-k_0)x_0}.
\eeq

Thus, (\ref{pack}) is finally rewritten as:
\beq
|t=0\rangle = \pin{k_1} e^{-\sigma^2(k_1-k_0)^2} e^{-i(k_1-k_0)x_0}|k_1\rangle. \label{pak1}
\eeq
Impose $x_0 \ll -1/m$ and $k_0 \gg 1/\sigma$, why?
\begin{itemize}
    \item For $x_0 \ll -1/m$: Start far to the left of the kink (kink centered at $x = 0$), so that initially the meson is well separated from the kink
    \item For $k_0 \gg 1/\sigma$: The width of the momentum distribution ($\Delta k  \sim 1/\sigma$) is small compared to $k_0$. This ensures that the wavepacket moves mostly in one direction, i.e., rightward.
\end{itemize}
It is important to note here that 
we create a wavepacket coming from the left moving to the right, thus, by construction, we do not have mesons coming in from the right. We only have mesons coming from the left that scatter off the kink.
To obtain the dynamics, we would let the packet evolve with $e^{-iE{(k_1)} t}$. Under time evolution using the Schrodinger picture,
\beq
|t\rangle = e^{-iHt}|t = 0\rangle = \pin{k_1} e^{-\sigma^2(k_1-k_0)^2} e^{-i(k_1-k_0)x_0} e^{-iE{(k_1)} t}|k_1\rangle. \label{evolve}
\eeq
The energy $E(k_1)$ is that of a meson moving in the kink background with momentum $k_1$.

Recall that 
\beq
E(k) = \omega_k = \sqrt{k^2 + m^2}
\eeq
Since the wavepacket is sharply peaked around $k = k_0$, we can expand $\omega(k_1)$ around $k_0$ using the Taylor expansion of first order
\bea
\omega(k_1) &\approx& {\omega(k_0) + \frac{\partial \omega(k)}{\partial k}}\Big|_{k_0} (k_1 - k_0)\\
\omega(k_1) &\approx& {\omega(k_0) + \frac{k_0}{\omega(k_0)}} (k_1 - k_0)\nonumber
\eea
then 
\beq
e^{-i E(k_1)t} \approx e^{-i\omega(k_0) t} \cdot e^{-i \frac{k_0}{\omega(k_0)} (k_1 - k_0)t}
\eeq
where the first term on the right-hand side is a constant phase and the second term is a linear phase in $k_1 - k_0$ which will shift the wave packet.

Eq.(\ref{evolve}) becomes 
\bea
|t\rangle &=& e^{-i \omega_{k_0} t}\pin{k_1} e^{-\sigma^2(k_1-k_0)^2} e^{-i(k_1-k_0)x_0} e^{-i \frac{k_0}{\omega_{k_0}} (k_1 - k_0)t}|k_1\rangle\\
&=& e^{-i \omega_{k_0} t}\pin{k_1} e^{-\sigma^2(k_1-k_0)^2} e^{-i(k_1-k_0)x_t}|k_1\rangle\nonumber
\eea
where $x_t = x_0 + \frac{k_0}{\omega_{k_0}}t$.
It is important to note that the wavepacket here represents a meson moving toward the kink with velocity $\frac{k_0}{\omega_{k_0}}$.

\underline{\underline{Digression}}:
The question here is, when this wavepacket moves toward the kink, what happens when it reaches the kink?
\begin{itemize}
    \item As the wavepacket hits the kink, it scatters - part of it may be reflected, part transmitted.
    \item 
    The scattering is characterized by both a reflection coefficient and a phase shift. These two quantities are independent; for example, the classical $\phi^4$ kink is reflectionless ($R(k) = 0$) but still exhibits a nonzero phase shift. The phase shift appears in the asymptotic behavior of the kink Hamiltonian eigenstate $|k_1\rangle$, which behaves as
    \bea
    |k_1\rangle &\sim&
        \begin{cases}
        e^{i k_1 x} + R(k_1)e^{-i k_1 x}, &  x \ll 0, \\
        T(k_1)e^{i k_1 x}, &  x \gg 0,
        \end{cases}
    \eea
\end{itemize}
for the reflection and transmission coefficients $R(k_1)$ and $T(k_1)$.

\subsubsection{Scattering Probability}
To calculate the scattering probability in kink-meson scattering or more generally in a quantum scattering problem involving solitons, mirrors standard quantum field theory scattering procedure, however, adapted to the non-perturbative background of the kink.\\
To achieve this scattering probability calculation, the 
procedure proceeds as follows:
\begin{enumerate}
    \item Define Initial WavePacket\\
    Construct an in-state wave packet of mesons in the presence of a kink (\ref{pak1}). This wavepacket is sharply peaked at momentum $k_0$, centered far to the left of the kink, and built from Hamiltonian eigenstates $|k_1\rangle$ that include mesons interacting with the kink.
    \item Evolve in Time\\
    Time evolve this state under full Hamiltonian
    \beq
    |t\rangle = \pin{k_1}e^{-\sigma^2(k_1 - k_0)^2}e^{-i(k_1 - k_0)x_0}e^{-i\omega_{k_1}t}|k_1\rangle
    \eeq
    \item Expand using Lippmann-Schwinger Form\\
    Insert the Lippmann-Schwinger form for $|k_1\rangle$ that contains the pole structure 
    \beq
    |k_1\rangle = \pin{k_2}\left[F(k_1, k_2) + \frac{R(k_1)}{k_1 + k_2 + i\epsilon}\right]|k_2\rangle_0.
    \eeq
    Plug this into the wavefunction and rearrange to obtain:
    \beq
    |t\rangle = \pin{k_2} I\left(k_2\right)|k_2\rangle_0
    \eeq
    where 
    \bea
    I(k_2) = \pin{k_1}e^{-\sigma^2(k_1 - k_0)^2}e^{-i(k_1 - k_0)x_t}\left[F(k_1, k_2) + \frac{R(k_1)}{k_1 + k_2 + i\epsilon}\right]\nonumber
    \eea
    \item Extract Final wavepacket: At late times\\
    When $x_t\gg0$ (i.e long after the meson passed the kink), one again deforms the contour, now enclosing the pole at $k_1 = -k_2$.\\
    Compute the residue of the pole giving\\
    \bea
    I_{pole}{(k_2)} \approx 2\pi i \cdot Res\left[\frac{R(k_1)}{k_1 + k_2 + i\epsilon} e^{....}\right]_{k_1 = -k_2}.\nonumber
    \eea
    This gives the elastic scattering component of the final wavepacket.\\
    \beq
    I_{pole}{(k_2)} = -i R(-k_2)e^{-\sigma^2(k_1 - k_0)^2}e^{-i(k_1 - k_0)x_t}
    \eeq
    The reflected part of the state is
    \beq
    |t\rangle_{refl} = -ie^{-i\omega_{k_0}t} \pin{k_2}R(-k_2)e^{-\sigma^2(k_1 - k_0)^2}e^{-i(k_1 - k_0)x_t}|k_2\rangle_0
    \eeq
    The coefficient of the final free meson state $|k_2\rangle_0$ tells the amplitude.\\
    The justification for evaluating the amplitude at the peak of the Gaussian is that the wavepacket is sharply localized in momentum space; therefore the integral is dominated by contributions near $k_2 \approx k_0$, yielding $R(k_0)$. This approximation is valid when $\sigma\gg\frac{1}{|k_0|}$ and $\sigma\gg\frac{1}{m}$.\\
    From here, the elastic scattering probability is
    \beq
    P(k_0) = |R(k_0)^2|.
    \eeq
\end{enumerate}

\section{Scattering Amplitude Example for The \texorpdfstring{$\Phi^4$}{Phi4} Model} \label{exsez}
\subsection{Generalized Scattering Amplitude Calculation}
The contributions of the generalized amplitudes are shown 
in Figure~\ref{phfig} below.

\begin{figure}[htbp]
\centering
\includegraphics[width = 1.0\textwidth]{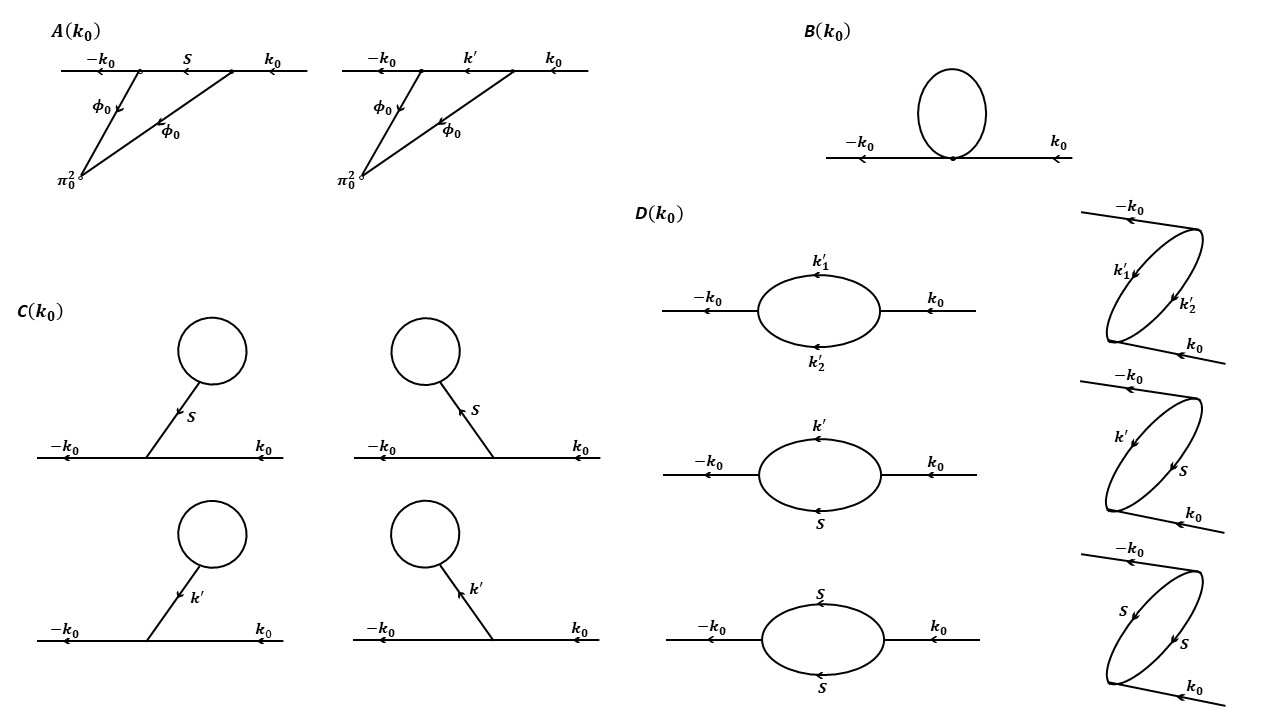}
\caption{The diagrams represent the contributions of individual amplitudes to the elastic scattering amplitude $R(k_0)$. Note that here time starts from the right.}
\label{phfig}
\end{figure}


The amplitude of the contribution is
\beq\label{AmplitudeDef}
R(k_0) = \lambda(A(k_0) + B(k_0) + C(k_0) + D(k_0))
\eeq

where
\begin{align} \label{scat}
A(k_0) &= \frac{1}{4 \lambda Q_0 k_0}\left[\sum_{S} \left(\frac{\omega_{k_0}^2 + \omega_{S}^2}{\omega_{S}}\right) \Delta_{-k_0-S} \Delta_{-k_0 S} + \pin{k^\prime}\left(\frac{\omega_{k_0}^2 + \omega_{k^\prime}^2}{\omega_{k^\prime}}\right)\Delta_{-k_0-k^\prime} \Delta_{-k_0 k^\prime}\right], \nonumber \\[1em]
B(k_0) &= \frac{V_{I -k_0 -k_0}}{4 k_0}, \nonumber \\[1em]
C(k_0) &= -\frac{1}{4 k_0}\left[\sum_{S} \frac{V_{IS} V_{-S -k_0 -k_0}}{\omega_{S}^2} + \pin{k^\prime} \frac{V_{I k^\prime}V_{-k^\prime -k_0 -k_0}}{{\omega_{k^\prime}^2}}\right], \nonumber \\[1em]
D(k_0) &= \frac{1}{8k_0}\int\frac{dk'_1dk'_2}{(2\pi)^2}\frac{(\omega_{k_1\p} + \omega_{k_2\p}) V_{-k_0 k_1\p k_2\p} V_{-k_0-k_1\p -k_2\p}}{\omega_{k_1\p} \omega_{k_2\p} \left(\omega_{k_0}^2 - \left(\omega_{k_1\p} + \omega_{k_2\p}\right)^2 + i \epsilon\right)} \\ 
&+ \frac{1}{8k_0}\pin{k^\prime}\frac{(\omega_{k'} + \omega_{S}) V_{-k_0 k'S} V_{-k_0-k'S}}{\omega_{k\p} \omega_{S} \left(\omega_{k_0}^2 - \left(\omega_{k\p} + \omega_{S}\right)^2 + i \epsilon\right)} +\frac{1}{4k_0}\sum_{S}\frac{V_{-k_0 S S}V_{-k_0 -S -S}}{\omega_S \left(  \omega^2_{k_0}- 4 \omega_{S}^2 + i\epsilon\right)}. \nonumber
\end{align}

From Eq.~(\ref{scat}), it is shown that we have the interaction vertex factor which corresponds to a local interaction involving the modes $k_i$, with internal loops that originate and terminate at the same vertex. These modes may represent a shape mode or a meson. However, zero modes behave differently: they do not form such loops. Instead, their contributions to interaction vertices are expressed directly in terms of the translation matrix elements $\Delta$, or equivalently, through their relation to lower-order vertices without zero-mode contributions.

These interaction vertices, together with their zero-mode reductions in terms of translation matrix elements $\Delta$, can be explicitly determined once the normal modes and the potential of the model are specified.
In Ref.~\cite{2023egm}, the necessary formulas are provided to determine these quantities.

In the $\phi^4$ double-well model, the small fluctuation spectrum consists of a translational mode, a continuum mode, and a single discrete shape mode, as given in Eq.~(\ref{modes}). Therefore, the internal line $k\p$ in the four processes shown in Figure 1 may also propagate via the shape-mode channel.

For process A, one may view the incident meson as gently imparting recoils to the kink's center of mass. After two such recoil insertions, the kink typically relaxes back to its ground state; along the way, brief excitations of both the continuum and the discrete shape mode can occur. 
Process B can be regarded as a four-point contact interaction, without explicit recoil insertions and the discrete shape-mode contributions.

In the process C, the reflected meson is accompanied by a loop that effectively annihilates a virtual meson left behind. The associated loop momentum or the continuum mode integral may also include contributions from the discrete shape mode.

The interesting part happens in process D. We can organize it as
\begin{equation}\label{DSplit}
    D(k_0)=D(k_0)_{k\p k\p}+D(k_0)_{S+k\p}+D(k_0)_{SS},\nonumber
\end{equation}
where $D_{k\p k\p}$ denotes the two-continuum mode integral, $D_{S+k\p}$ mixes one shape mode with one continuum mode, and $D_{SS}$ corresponds to the two-shape mode excitation. The terms $D_{k\p k\p}$ and $D_{S+k\p}$  set the smooth background and thresholds for multi-meson production. By contrast, when the incoming meson energy approaches twice the shape-mode energy, $\omega_{k_0}^2 \approx 4\omega_{S}^2$, the denominator can develop a pole. After Dyson resummation (bubble dressing) of the two-shape mode propagator, this pole is expected to shift into the complex plane, leading to a Breit-Wigner type resonance with a finite width and lifetime.

Now that our focus is on the $\phi^4$ model, the specified potential to be used is important for us to be able to compute the relevant functions or quantities stated in Eq.~(\ref{scat}).

\subsection{\texorpdfstring{$\Phi^4$}{Phi4} Model Example}
Looking into $\phi^4$ double well theory, whose potential is considered 
\beq
V(\sqrt{\lambda} \phi(x))=\frac{ \lambda \phi(x)^2}{4} \left(\sqrt{\lambda}\phi(x) - m \sqrt{2}\right)^2
\eeq 
the stationary kink classical solution is
\beq
\phi(x) = f(x) = \frac{m}{\sqrt{2\lambda}}\left[\tanh\left[\frac{m}{2}x\right] + 1 \right].
\eeq
The classical kink solution's 
mass is
\beq
Q_0 = \frac{m^3}{3 \lambda}.
\eeq
The continuum normal modes, shape mode, and zero mode of the $\phi^4$ kink are 
\bea\label{modes}
\g_k(x)=\frac{2 e^{-i k x}}{\omega_k \sqrt{m^2 + 4 k^2}}\left[k^2 - \frac{m^2}{2} + \frac{3 m^2}{4}\sech^2\left(\frac{m x}{2}\right) - i \frac{3 m}{2}k\tanh\left(\frac{m x}{2}\right)\right] \\
\g_{S}(x)=\frac{3\sqrt{m}}{2\sqrt{3}} \tanh\left(\frac{m x}{2}\right)\sech\left(\frac{m x}{2}\right)\hsp
\g_B(x)=-\frac{\sqrt{3 m}}{2\sqrt{2}}\sech^2\left(\frac{m x}{2}\right).\nonumber
\eea
which have frequencies 
\beq
\omega_k = \sqrt{m^2 + k^2} \hsp
\omega_{S} = \sqrt{3}\frac{m}{2}  \hsp
\omega_B = 0.
\eeq

\subsubsection{Relevant functions computation}
We can find the loop factor and the matrix using Ref.~\cite{2022xdz}
\bea
\I(x)&=&\pin{k}\frac{\left|{\g}_{k}(x)\right|^2-1}{2\omega_k}+\sum_S \frac{\left|{\g}_{S}(x)\right|^2}{2\omega_{S}}, \\
\Delta_{i j}&=&\int dx \g_{i}(x) \g\p_{j}(x),\nonumber
\eea
which gives
\beq
\I(x) = \frac{\sech^2\left[\frac{m x}{2}\right]\tanh^2\left[\frac{m x}{2}\right]}{4\sqrt{3}} - \frac{3\sech^4\left[\frac{m x}{2}\right]}{8\pi},
\eeq

\bea
\Delta_{k_1 k_2} &=& i\pi(k_1 - k_2)\delta(k_1 + k_2) + i\pi3\left(\frac{\omega_{k_1}}{\omega_{k_2}} - \frac{\omega_{k_2}}{\omega_{k_1}}\right) \\
&& \times \frac{k_1^2 + k_2^2 + m^2}{\sqrt{m^2 + 4 k_1^2}\sqrt{m^2 + 4 k_2^2}}\csch\left[\frac{(k_1 + k_2) \pi}{m}\right], \nonumber \\ [1em]
\Delta_{k S} &=& \frac{\pi\sqrt{3}\left(16 k^4 + 16 m^2 k^2 + 3 m^4\right)}{16 m^{3/2} \omega_{k}\sqrt{m^2 + 4 k^2}}\sech\left[\frac{k \pi}{m}\right]. \nonumber
\eea

Using the soliton solution, the potential derivatives are easily found
\bea
V^{(3)}\left(\sqrt{\lambda} f(x)\right) = 3 m\sqrt{2 \lambda} \tanh\left[\frac{mx}{2}\right]\hsp
V^{(4)}\left(\sqrt{\lambda} f(x)\right) = 6 \lambda.
\eea

Now that we have these, the n-point interactions $V$ can be evaluated. 
Based on what we need from Eq(3.1), we will first pay attention to the interaction $V$ that contains a loop.
\bea
V_{\I k} &=& i\frac{k^2 \omega_k \sqrt{\lambda}}{m^4 \sqrt{6}\sqrt{m^2 + 4k^2}}\left[\pi(2k^2 - m^2) + 3\sqrt{3}\omega_k^2\right]\csch\left[\frac{k \pi}{m}\right], \\
V_{\I S} &=& -\frac{3 \sqrt{m \lambda}}{128 \sqrt{2}} \left(-2\pi + 3\sqrt{3}\right). \nonumber
\eea

The interactions, which include the shape mode, remain finite and constitute independent contributions
\bea
 V_{k_1 k_2 S} &=& \frac{\pi 3\sqrt{3 \lambda}}{2\sqrt{2} m^{3/2} \omega_{k_1} \omega_{k_2} \sqrt{m^2 + 4k_1^{2}} \sqrt{m^2 + 4 k_2^{2}}} \\
&& \times \left[8 m^2 k_1^2 k_2^2   +   \left(m^2 + 4 k_1^2 + 4 k_2^2\right) \left(\frac{17 m^4}{16} - \left(k_1^2 - k_2^2\right)^2\right) \right]\sech \left[\frac{ \left(k_1 + k_2\right)\pi}{m}\right] \nonumber \label{VSk1k2}.
\eea


However, the three-meson coupling diverges when the sum of the external momenta vanishes, corresponding to zero momentum exchange with the kink. The divergent component of $V_{k_1 k_2 k_3}$ arises from the vacuum energy and must therefore be subtracted in the two-loop kink-mass calculation. After this subtraction, only the finite part remains, which represents the genuine physical vertex correction contributing to the scattering amplitude Ref.~\cite{2021gxs}:

\begin{equation}\label{vk1k2k3}
    \begin{split}
        V_{k_1k_2k_3} = & \frac{-i \pi \sqrt{2 \lambda}}{\ok{1} \ok{2} \ok{3}\sqrt{(4\omega^2_{k_1} - 3m^2)(4\omega^2_{k_2} - 3m^2)(4\omega^2_{k_3} - 3m^2)}}\\ 
        & \times\left(N_0(k_1,k_2,k_3) \delta(k_1+k_2+k_3) + N_1(k_1,k_2,k_3) \csch\left(\frac{(k_1+k_2+k_3)\pi}{m}\right)\right),
    \end{split}
\end{equation}
where
\begin{align}
        N_0(k_1,k_2,k_3) &= 18m^2\Biggl[m^4(k_1+k_2+k_3)+4k_1k_2k_3(k_1(k_2+k_3)+k_2k_3) \\ 
        &-m^2(2k^2_1(k_2+k_3)+k_1(2k^2_2+9k_2k_3+2k^2_3)+2k_2k_3(k_2+k_3)) \Biggr],\nonumber \\
        N_1(k_1,k_2,k_3)  &= 3\Biggl[3k^6_1 - 3k^4_1(k^2_2 + k^2_3) + 3(k^2_2 - k^2_3)^2(k^2_2 + k^2_3) - 8k^2_2k^2_3m^2 \\
        & -5(k^2_2+k^2_3)m^4-2m^6- k^2_1(3k^4_2+ 3k^4_3+2k^2_2k^2_3+8(k^2_2+k^2_3)m^2+5m^4)\Biggr].\nonumber
\end{align}

We also need the
four-point coupling to compute the leading-order elastic scattering and the shape mode-meson coupling
\begin{align}
        V_{Ikk}=&\frac{4 k(4k^2(9+2\sqrt{3}\pi)+m^2(11\sqrt{3}\pi-45))}{105m^4 \lambda}\csch\left(\frac{2k\pi}{m}\right), \label{VIKK1} \\
       V_{kSS}=&-i\pi\frac{{3\sqrt{\lambda}}}{\sqrt{2}}\frac{k^2\ok{}(m^2-2k^2)}{m^3\sqrt{m^2+4k^2}}.
\end{align}

\subsubsection{Numerical Calculation of the Contributions in the \texorpdfstring{$\Phi^4$}{Phi4} model}
Substituting these into our general result Eq.~(\ref{scat}), we find the individual contributions to soliton-meson scattering in the $\Phi^4$ model, which has shape modes unlike the sine-Gordon model in Ref.~\cite{2023egm}
\allowdisplaybreaks
\begin{align}
A(k_0) 
&= \frac{9 \pi^2}{2^{10} m^6 k_0\ok {0}^2}\frac{\left(\ok{0}^2 + \omega_S^2\right)\left(4\ok {0}^2 - m^2\right)^2 \left(4 \ok{0}^2 - 3 m^2\right)}{\omega_S} \sech^2\left(\frac{\pi k_0}{m}\right) \nonumber \\
&\quad - \frac{27 \pi^2}{4 m^3 k_0\ok {0}^2 \left(4\ok{0}^2 - 3m^2\right)}\pin{k\p}\frac{(\ok 0^2+\okp{}^2)(\ok 0^2-\okp{}^2)^2(\ok 0^2 + \okp{}^2 - m^2)^2}{\okp{}^3 (4 \okp{}^2 - 3 m^2)} \nonumber \\
&\quad \times \csch\left( \frac{\pi(k_0+k\p)}{m}\right)\csch\left( \frac{\pi(k_0-k\p)}{m}\right), \\[1em]
B(k_0)
&= \frac{ \lambda \big(4(9 + 2\sqrt{3}\pi)\ok 0^2+3(\sqrt{3}\pi - 27) m^2\big)}{105 m^4}
\csch\!\left(\frac{2\pi k_0}{m}\right), \\[1em]
C(k_0) \label{ck0}
&= \frac{9 \sqrt{3}\pi m \lambda}{2^{14} k_0\ok 0^2 (m^2 + 4 k_0^2)}\frac{(128 k_0^4 + 17(m^4 + 8m^2 k_0^2))(-2\pi + 3\sqrt{3})}{\omega_S^2}\sech\left(\frac{2 \pi k_0}{m}\right) \nonumber \\
&\quad - \frac{18 \sqrt{3} k_0^4 \lambda}{ m^2 \ok{0}^2 \omega_{2k_0}^2 (m^2 + 16 k_0^2)} \left[\omega_{2k_0}^2 (3 \pi  + 3\sqrt{3}) - 4 \pi \ok{0}^2\right] \csch\left(\frac{2\pi k_0}{m}\right) \nonumber \\
&\qquad  + \frac{\pi \lambda}{4 m^4 k_0 \ok{0}^2 \omega_{2k_0}^2} \pin{k\p}\frac{{k\p}^2 \left[(2 \sqrt{3}\pi  + 9) \omega_{k'}^2  -  3\sqrt{3}\pi m^2\right] \left[3{k\p}^4 - 3(k_0^2 + \ok{0}^2){k\p}^2 - 2 \ok{0}^2 \omega_{2k_0}^2 \right]}{\omega_{2k'}^2} \nonumber \\
&\qquad \times \csch\left( \frac{\pi k\p}{m}\right)\csch\left(\frac{\pi(k\p + 2k_0)}{m}\right), \\[1em]
D(k_0) \label{dk0}
&= -\frac{9 k^3_0\pi^2\lambda\omega^2_{k_0}(3m^2 - 2\omega^2_{k_0})^2}{8m^6\omega_S(4\omega^2_{k_0}-3m^2)(\omega^2_{k_0}-4\omega^2_S+i\epsilon)}\csch^2\left(\frac{\pi k_0}{m}\right) \nonumber \\
&\quad + \frac{27 \pi^2\lambda}{2^{14} k_0 m^3\omega^2_{k_0}\omega_S(4\omega^2_{k_0}-3m^2)} \nonumber \\
&\qquad \times \int\frac{dk'}{2\pi}\frac{(\omega_{k'} + \omega_S)[m^2 (17 m^4  +  68 m^2 k_{+}  +  32 k^2_{+}) - 16 k^2_{-} (3 m^2  +  4 k_{+})]^2}{\omega^3_{k'}(4\omega^2_{k'}-3m^2)(\omega^2_{k_0}-(\omega_{k'} + \omega_S)^2+i\epsilon)} \nonumber \\
&\qquad \times \sech\left(\frac{(k_0 - k')\pi}{m}\right)\sech\left(\frac{(k_0 + k')\pi}{m}\right) \nonumber \\
&\quad + F_{10}(k_0)\int\frac{dk'_1}{2\pi}F_{11}(k_0,k'_1)\left(Q_{-}(k_0,k'_1)-Q_{+}(k_0,k'_1)\right) \nonumber \\
&\quad + F_{21}(k_0)\int\frac{dk'_1 dk'_2}{(2\pi)^2}F_{22}(k_0,k'_1,k'_2).
\end{align}

Where the functions $F_{10}(k_0),F_{11}(k_0,k'_1)$ and $Q_{\mp}(k_0,k'_1)$ are:
\begin{equation}
    \begin{split}
        F_{10}(k_0) = & \frac{27m^2\pi\lambda}{2 \ok{0}^2 (4\omega^2_{k_0}-3m^2)}\csch\left(\frac{2k_0\pi}{m}\right), \\ 
        F_{11}(k_0,k'_1) = & \frac{k'_1 }{(4\omega^2_{k'_1}-3m^2)\omega^3_{k'_1}}, \\
        Q_{\mp}(k_0,k'_1)= & \frac{F_{\mp}(k_0,k'_1)P_{\mp}(k_0,k'_1)}{D_{\mp}(k_0,k'_1)},
    \end{split}
\end{equation}
with
\begin{equation}
    \begin{split}
        F_{\mp}(k_0,k'_1)=&(k_0\mp k'_1)(4(k^2_0\mp k_0k'_1+{k'_1}^{2})+3m^2), \\ 
        P_{\mp}(k_0,k'_1)=&(S_{\mp}(k_0,k'_1)+L_{\mp}(k_0,k'_1))(\omega_{k'_1} + \omega_{k_0\mp k'_1}), \\ 
        S_{\mp}(k_0,k'_1)=& 8k^2_0(k_0\mp k'_1)^2{k'_1}^2-4(k^2_0\mp k_0k'_1+{k'_1}^2)^2m^2, \\
        L_{\mp}(k_0,k'_1)=& -5(k^2_0\mp k_0k'_1 +{k'_1}^2)m^4-m^6, \\
        D_{\mp}(k_0,k'_1)=&({4\omega^2_{{k_0\mp k'_1}}} -3 m^2)\omega^3_{{k_0\mp k'_1}}(\omega^2_{k_0} - (\omega_{k'_1} + \omega_{{k_0\mp k'_1}})^2 + i\epsilon).
    \end{split}
\end{equation}

The functions $F_{21}(k_0)$ and $F_{22}(k_0,k'_1,k'_2)$ defined in the 2D integration are as follows

\begin{equation}
    \begin{split}
        F_{21}(k_0)=&\frac{-9\pi^2\lambda}{4k_0 \ok{0}^2 (4\omega^2_{k_0}-3m^2)},\\
        F_{22}(k_0,k'_1,k'_2)=&\frac{[f(k_0,k'_1,k'_2)]^2 (\omega_{k'_1}+\omega_{k'_2})}{D_0(k'_1,k'_2)D_1(k'_1)D_2(k'_2)D_3(k_0,k'_1,k'_2)}\\
        & \times\csch\left(\frac{(k_0 - k'_1 - k'_2)\pi}{m}\right)\csch\left(\frac{(k_0 + k'_1 + k'_2)\pi}{m}\right).
    \end{split}
\end{equation}
 The denominators are 
 \begin{equation}
     \begin{split}
         D_0(k'_1,k'_2)=& \omega^3_{k'_1}\omega^3_{k'_2}, \quad
         D_{1,2}(k'_{1,2})= (4\omega^2_{k'_{1,2}}-3m^2), \quad 
         D_3(k_0,k'_1,k'_2)=(\omega^2_{k_0}-(\omega_{k'_1}+\omega_{k'_2})^2+i\epsilon)
     \end{split}
 \end{equation}
and the numerator function $f(k_0,k'_1,k'_2)$ is 
\begin{equation}
    \begin{split}
        f(k_0,k'_1,k'_2)=&3a^3 - 3a^2s - a(3s^2 - 4p + 8sm^2 + 5m^4) + 3s^3 - 12sp - 8pm^2 - 5sm^4 - 2m^6
    \end{split}
\end{equation}
where the coefficients are 
\begin{equation}
            a=k^2_0, \quad s={k'_1}^2+{k'_2}^2, \quad p={k'_1}^2{k'_2}^2, \quad k_{\pm}=k^2_0\pm {k'}^2.
\end{equation}

\newpage
\begin{figure}[htbp]
\centering
\includegraphics[width = 0.95\textwidth]{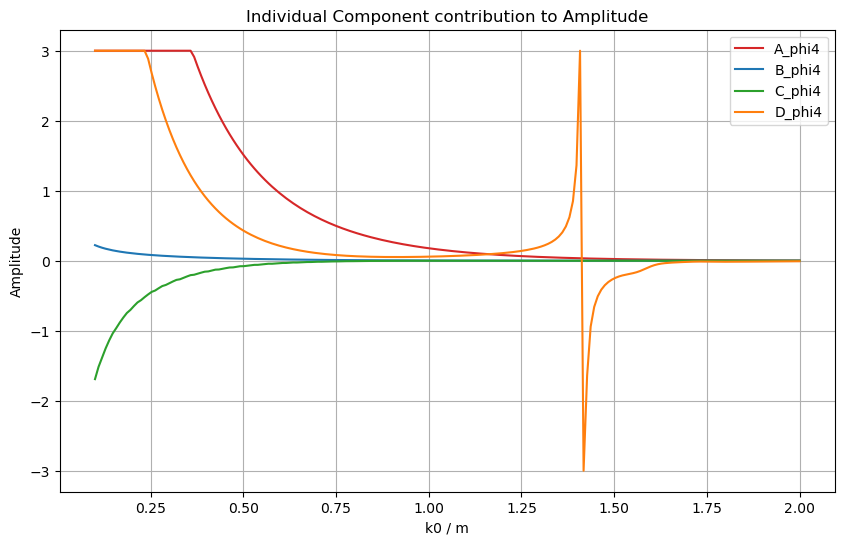}
\caption{Contributions $A(k_0)$, $B(k_0)$, $C(k_0)$ and $D(k_0)$ to the elastic scattering in the $\phi^4$ model}\label{sgfig}
\end{figure}

\begin{figure}[htbp]
\centering
\includegraphics[width = 0.95\textwidth]{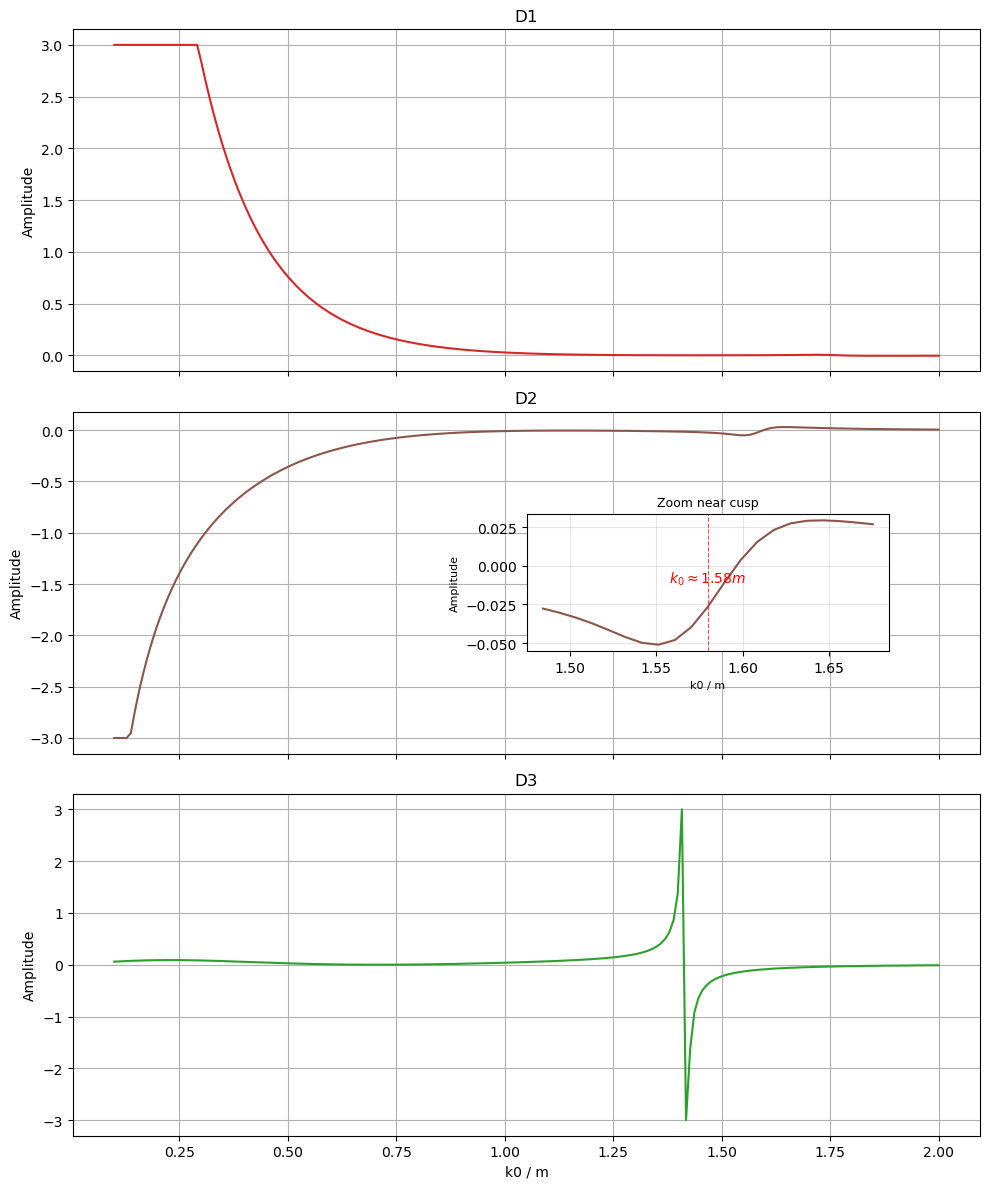}
\caption{Decomposition of the $D(k_0)$ contribution into two-meson continuum (D1), meson–shape-mode continuum (D2), and two-shape-mode pole (D3) channels. The cusp-like feature near $k_0 \approx 1.58m$ appears only in D2, confirming its origin in the meson–shape-mode threshold.}\label{sgfig3}
\end{figure}

\newpage
\section{Remarks} 
In this work, we have applied the framework of Linearized Soliton Perturbation Theory 
(LSPT) to compute the one-loop elastic scattering amplitude of a meson off the $\phi^4$ kink. As in the general analysis of Ref.~\cite{2023egm}, the full amplitude decomposes into four contributions, $A(k_0), B(k_0), C(k_0),$ and $D(k_0)$, corresponding to the four diagrammatic processes in Fig.~\ref{sgfig}. Because the $\phi^4$ kink supports a single bound shape mode, these contributions combine into a structure with several noteworthy physical features.

The most prominent qualitative feature is the sharp peak in the $D(k_0)$ contribution located at $k_0 \simeq \sqrt{2}m$ arising from the denominator of the \emph{two-shape-mode} intermediate-state propagator. At this energy, which satisfies 
$\omega_0 \approx 2\omega_S$, the incoming meson is resonant with a twice-excited shape-mode configuration, producing a pole in the scattering amplitude.
As discussed in Sec. 3 of Ref.~\cite{2023egm}, this intermediate configuration is unstable in the $\phi^4$ model. Consequently, the pole visible at leading order is expected to broaden into a Breit–Wigner type resonance once repeated two-shape-mode bubbles are resummed. This is closely analogous to resonant structures
in nuclear and hadronic scattering, where near-on-shell intermediate states generate rapid energy dependence in the elastic amplitude\cite{1958nx}.



For larger external momentum, $k_0 \gtrsim 1.6m$, the amplitude develops nontrivial analytic structure associated with continuum thresholds. In particular, the continuum-shape channel exhibits a branch cut when the incoming energy satisfies $\omega_0 = \omega_k + \omega_S$, reflecting the onset of a meson-shape-mode continuum state. Similarly, the continuum-continuum channel develops branch cuts at $\omega_0 = 2\omega_k$, corresponding to the two-meson threshold. These threshold effects can be analytically continued into the complex plane using the standard $i\epsilon$ prescription, where they modify the real part of the amplitude through dispersive corrections and produce sharp, cusp-like structures (sudden bends at threshold energies) in the elastic scattering amplitude\cite{1962pb}.

A particularly clear example of this behaviour occurs near $k_0 \approx 1.58 m$, where the amplitude displays a cusp-like feature associated with the opening of the meson–shape–mode continuum. This nonanalyticity originates from the second term in $D(k_0)$, whose energy denominator becomes singular when the incoming on-shell energy matches that of an on-shell intermediate state, $\omega_{k_0} = \omega_{k'} + \omega_S$. The threshold is determined by the lowest-energy configuration of this intermediate state. Since the continuum meson dispersion $\omega_{k'}$ is minimized at $k' = 0$, the first channel to open corresponds to a meson at rest with $\omega_{k'=0} = m$, implying $\omega_{k_0} = m + \omega_S$. Using the dispersion relation $\omega_{k_0} = \sqrt{k_0^2 + m}$, the corresponding threshold momentum is $k_0 = \sqrt{\omega_S^2 + 2 m \omega_S} \approx 1.58m$, in excellent agreement with the location of the feature observed in our numerical evaluation of $D(k_0)$. As expected, this nonanalytic structure arises solely from the meson–shape–mode intermediate state and does not appear in the other components of the amplitude.



To further verify the origin of the cusp‑like behavior, we have decomposed $D(k_0)$ into  its three physically distinct contributions:
\begin{itemize}
    \item the two-meson continuum channel (D1),
    \item the meson–shape–mode continuum channel (D2),
    \item the two–shape–mode pole channel (D3).
\end{itemize}
As shown in Fig.~\ref{sgfig3}, the cusp‑like nonanalytic structure appears exclusively in the $D2$ component, confirming that it is generated by the meson–shape–mode intermediate state. This supports the interpretation that the branch point at $k_0 \approx 1.58m$ marks the onset of the corresponding continuum threshold. In contrast, the other components remain smooth throughout this momentum region, reinforcing that the feature originates specifically from the meson–shape–mode channel.

Although the $\phi^4$ model contains neither spin nor isospin and possesses only a single mesonic degree of freedom, the kink acts as an extended, composite object-much like a baryon or light nucleus in effective descriptions. Scattering from such an extended background naturally exhibits resonant enhancement and energy-dependent nonlocality (arising from continuum dressing and threshold effects). These are the same analytic mechanisms that underlie a wide class of nuclear scattering phenomena, from $\pi N$ resonances to optical-model dispersive corrections. Closely related behavior also appears in soliton-based descriptions of baryons in the Skyrme model, where meson–baryon scattering displays resonances and nontrivial continuum dressing\cite{1983ya, 2004tk}.

Finally, unlike the integrable sine–Gordon model discussed in Ref.~\cite{2023egm}, the $\phi^4$ theory is non-integrable, and therefore no cancellation of the total amplitude occurs; that is, the four contributions do not sum to zero. Instead, they assemble into a finite, strongly energy-dependent scattering amplitude with both a narrow resonance and a smooth high-momentum tail. A more complete analysis, including higher-order resummations of the two-shape-mode channel and direct time-evolution simulations of wave-packet scattering, would further provide additional insight into both the width of the resonance and the analytic structure of the continuum contributions.



 
\section* {Acknowledgement}

\noindent
KO and BB thank Jarah Evslin for discussions crucial to the completion of this project.
KO was supported by the ANSO scholarship and the UCAS Partial Scholarship for International Students as a full-time Senior Visiting Student. Current support is provided by the Xi’an City Government Belt and Road Scholarship. BB is supported by the ANSO Scholarship CAS.
KO is also supported by the Higher Education and Science Committee of Armenia (HESCS), grant No. 25IRF/2-1C008

\end{document}